\documentclass[aps,pra,groupedaddress,superscriptaddress,showpacs,notitlepage]{revtex4-1}

\makeatletter
\renewcommand{\p@subsection}{}
\makeatother

\usepackage{amsfonts,amsmath,graphicx,dcolumn,bm,array,color,hyperref}
\usepackage{amssymb,setspace,empheq,braket,multirow,fixltx2e}
\usepackage[utf8]{inputenc}
\usepackage[english]{babel}
\usepackage[dvipsnames]{xcolor}
\usepackage{physics}
\usepackage{setspace}
\begin{document}
\raggedbottom

\title{Generation of optical Fock and W states with\\single-atom-based bright quantum scissors}

\author{Ziv Aqua}
\affiliation{AMOS and Department of Chemical and Biological Physics, Weizmann Institute of Science, Rehovot 76100, Israel}
\author{M. S. Kim}
\affiliation{QOLS, Blackett Laboratory, Imperial College London, London SW7 2AZ, United Kingdom}
\author{Barak Dayan}
\email{barak.dayan@weizmann.ac.il}
\affiliation{AMOS and Department of Chemical and Biological Physics, Weizmann Institute of Science, Rehovot 76100, Israel}

\date{\today}

\begin{abstract}
We introduce a multi-step protocol for optical quantum state engineering that performs as deterministic "bright quantum scissors" (BQS), namely truncates an arbitrary input quantum state to have \textit{at least} a certain number of photons. 
The protocol exploits single-photon pulses and is based on the effect of single-photon Raman interaction, which is implemented with a single three-level $\Lambda$ system (e.g. a single atom) Purcell-enhanced by a single-sided cavity. 
A single step of the protocol realises the inverse of the bosonic annihilation operator. Multiple iterations of the protocol can be used to deterministically generate a chain of single-photons in a W state. Alternatively, upon appropriate heralding, the protocol can be used to generate Fock-state optical pulses. This protocol could serve as a useful and versatile building block for the generation of advanced optical quantum states that are vital for quantum communication, distributed quantum information processing, and all-optical quantum computing.
\end{abstract}
\maketitle

\section{\label{sec:intro}Introduction}
The field of quantum state engineering (QSE) aims at preparing arbitrary quantum states. Nonclassical states are highly
sought after both as a means to test fundamental questions in quantum
mechanics \cite{pan2000experimental}, as well as a source for various applications in quantum
information \cite{bennett1993teleporting,jeong2001quantum}, sensing and metrology \cite{abadie2011gravitational}. Controlling and manipulating the
quantum state of optical fields is of particular interest both for optical information processing \cite{knill2001scheme,walther2005experimental} and for quantum
communication \cite{azuma2015all} since optical photons are the ideal carriers of information over long distances. There are two main approaches to engineer the
quantum state of an optical field \cite{dell2006multiphoton}: first, by choosing the Hamiltonian correctly,
one can utilise its time evolution to unitarily transform an initial
state into the desired final state (e.g. generation of squeezed states and entangeled photon pairs by parametric down-conversion). Second, by introducing entanglement between the system of interest and an auxilary system folloed by appropriate measurements on the auxilary system, one can collapse the system of interest to the target state. This approach was used for example for the generation and entanglement of single photons in the DLCZ protocol for long-distance quantum communication \cite{duan2001long}, and in the recent generation of entangled atom-light Schr\"{o}dinger cat states \cite{hacker2019deterministic}. The two approaches may of course be combined for instance in the generation of optical Schr\"{o}dinger cat states from squeezed vacuum, which is conditioned on the measurement of a subtracted photon diverted to an auxilary mode \cite{ourjoumtsev2006generating}. QSE of optical fields was discussed by Vogel et al. \cite{vogel1993quantum} in a paper proposing a recipe for generating an arbitrary quantum state
in the field of a single-mode resonator. Following that,
there have been considerable efforts on QSE of a traveling light field; from schemes preparing arbitrary quantum states using conditional measurements on beam splitters \cite{dakna1999generation,fiuravsek2005conditional},
to generating nonclassical states of specific interests such as single-photon
Fock states \cite{lvovsky2001quantum}, Schr\"{o}dinger cat states \cite{ourjoumtsev2006generating,ourjoumtsev2007generation}, NOON states \cite{afek2010high}, GHZ states \cite{bouwmeester1999observation,hamel2014direct}
and cluster states \cite{nielsen2004optical}. Moreover, many different manipulations of the quantum field were realised such as the annihilation and creation operators \cite{wenger2004non,zavatta2004quantum,parigi2007probing}, squeezing \cite{walls1983squeezed} and quantum scissors \cite{pegg1998optical}.

\begin{figure}[t]
\begin{center}
\includegraphics[width=12cm,trim={0 4cm 0 4cm},clip]{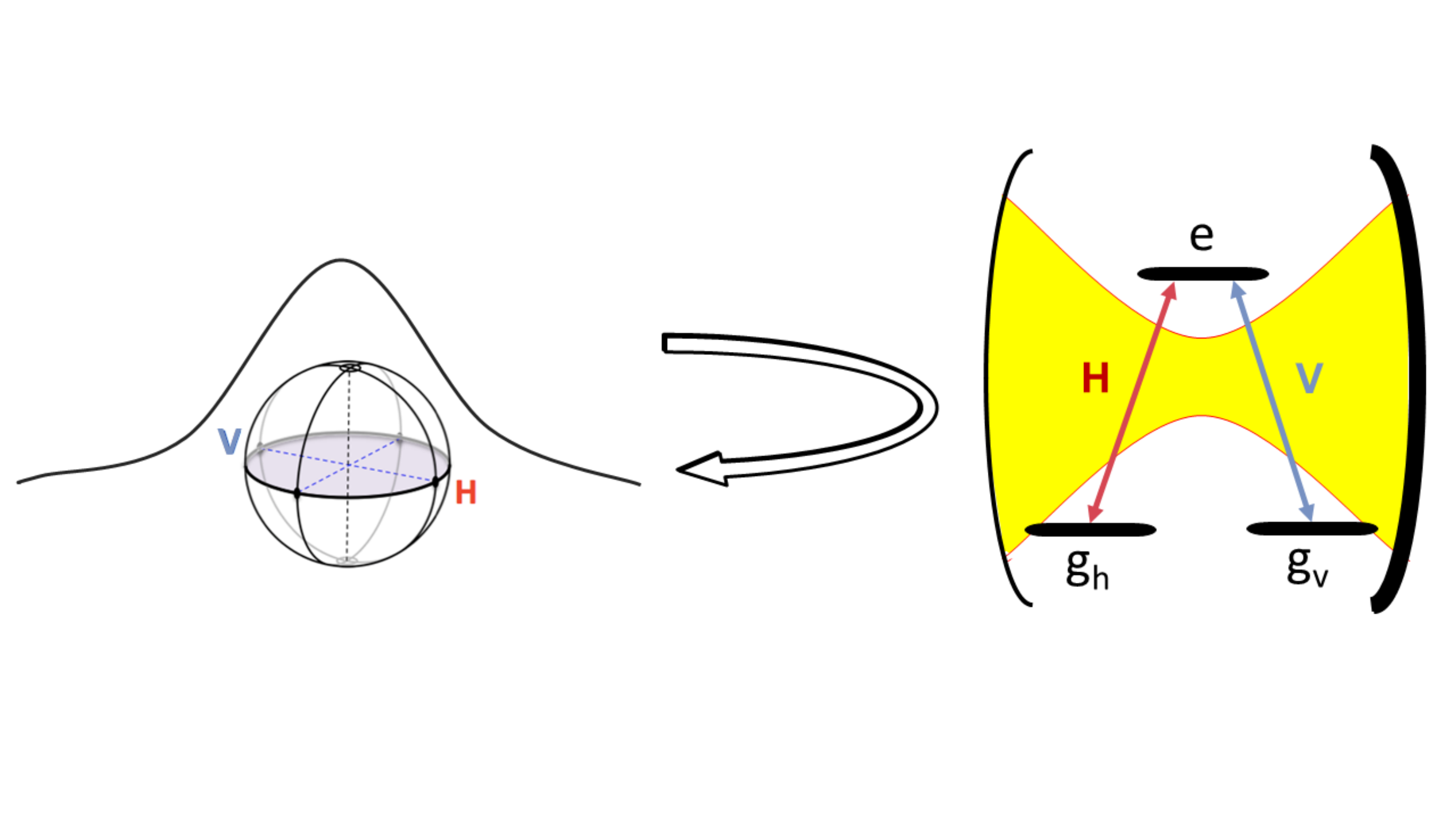}
\end{center}
\caption{The configuration that leads to single-photon Raman interaction (SPRINT). Two optical modes, in this case orthogonal polarisation (H and V), interacting with a three-level $\Lambda$ system in
a single-sided cavity. Each polarisation is coupled to a different
'leg' of the $\Lambda$ system. Upon an incident H-polarised single-photon pulse and a $\Lambda$ system prepared in $\ket*{g_h}$, destructive interference forces the $\Lambda$ system to emit back the photon in V and undergo
a Raman transition to state $\ket*{g_v}$. In effect this configuration realises a unitary swap gate between the photonic and atomic qubit.}
\label{Fig_SPRINT}
\end{figure}

\begin{figure}[h!]
	\centering\includegraphics[width=12cm]{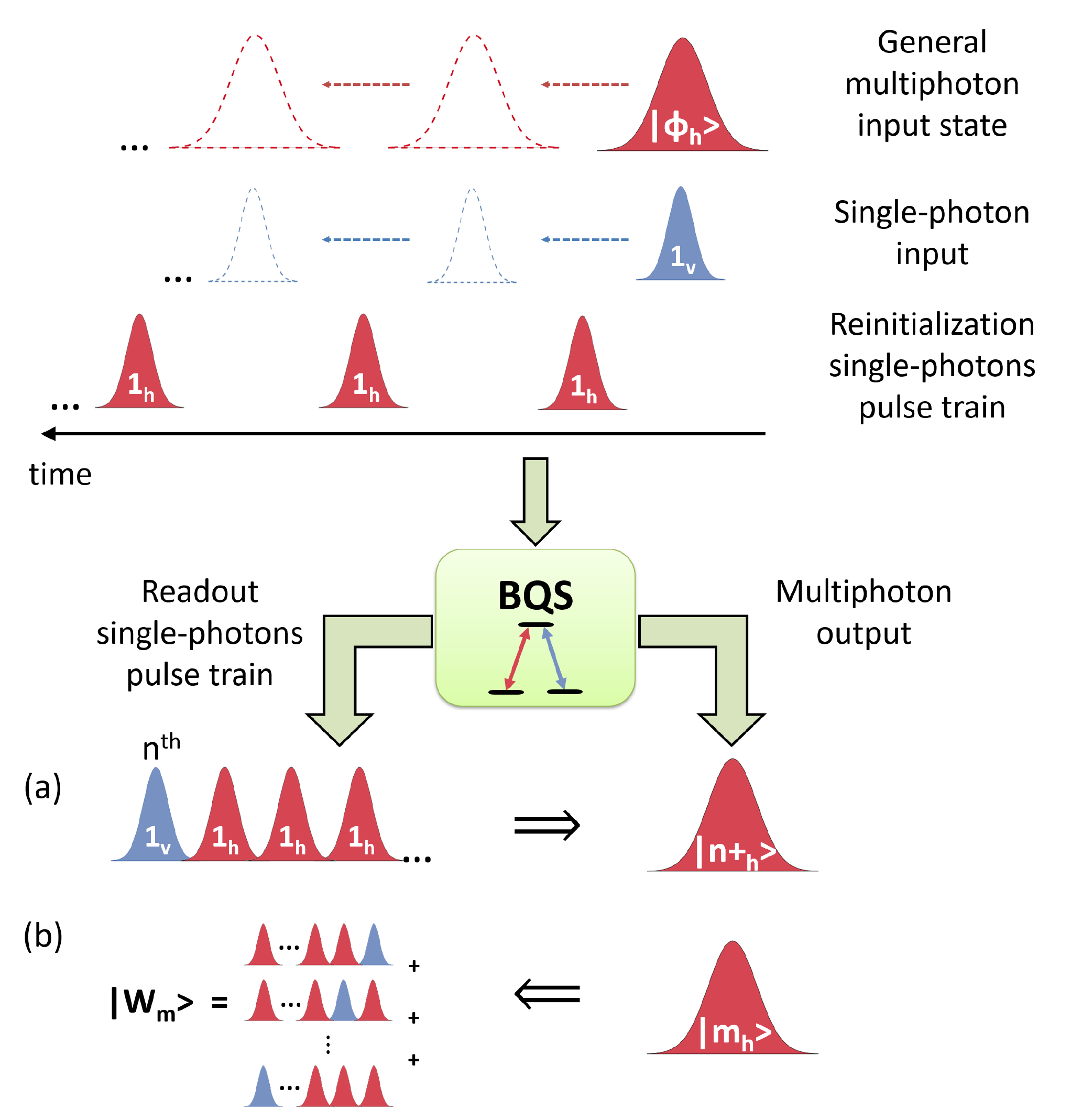}
	\caption{Bright quantum scissors (BQS) multi-step protocol. The protocol uses three input channels; a general H-polarised multiphoton quantum state, a V-polarised single photon and a train of H-polarised single photons. The multiphoton pulse and the V-polarised single-photon pulse interact with the $\Lambda$ system simultaneously and the resulting pulses are fed back to the system repeatedly.
	The H-polarised single-photon pulses are interleaved with the multiphoton pulse evolutions and reinitialize the state of the $\Lambda$ system at every iteration. At the output channels of the protocol we get a train of readout single-photon pulses and a modified multiphoton state. (a) Heralded on the measurement of the $n^{th}$ readout photon in the V-mode, the $n^{th}$-order BQS operation is applied on the input quantum state. This ensures the presence of more than $n$ photons in the multiphoton output. For $n=1$, the operation amounts to a realisation of the inverse annihilation. (b) Conversly, when choosing to herald on the number of photons in the multiphoton output pulse, a polarisation W-state manifests in the readout single-photons pulse train.}
	\label{Fig_BQS}
\end{figure}

At the heart of the study in this paper stands the single-photon
Raman interaction (SPRINT) \cite{rosenblum2016extraction,rosenblum2017analysis,bechler2018passive}. The configuration that leads to SPRINT was originally considered by Pinotsi and Imamoglu \cite{pinotsi2008single}
as an ideal absorber of a single photon. It was later studied in a series of theoretical works \cite{lin2009heralded,koshino2010deterministic,bradford2012single,koshino2017theory,rosenblum2017analysis} and shown to perform as a photon-atom swap gate and accordingly serve as a quantum memory. It was experimentally demonstrated with a single-atom coupled to a whispering-gallery mode (WGM) resonator and used to implement a single-photon router \cite{shomroni2014all}, extraction of a single photon from a pulse \cite{rosenblum2016extraction} and
a photon-atom qubit swap gate \cite{bechler2018passive}. In superconducting circuits it was demonstrated as well \cite{inomata2014microwave} and used for highly efficient detection of single microwave photons \cite{inomata2016single}. The SPRINT mechanism occurs in a three-level $\Lambda$ system where each transition is coupled to a single optical mode as shown in Fig. \ref{Fig_SPRINT} for the case of orthogonal polarisations H and V. As explained in detail in \cite{rosenblum2017analysis}, in this configuration a single H (V) photon is enough to send the atom to the corresponding dark state $\ket*{g_v}$ ($\ket*{g_h}$). Symmetrically, the polarization of the returning photon is set by the initial state of the atom - which makes this configuration perform as a photon-atom swap gate \cite{bechler2018passive}.
In this work, we explore the potential of the SPRINT mechanism in multi-photon processes within the theoretical framework of the ''modes of the universe'' (MOU) \cite{lang1973laser,gea1990treatment}. Specifically, we show that a single SPRINT-based iteration involving an arbitrary input quantum state in one optical mode and a single-photon pulse in the other can realise the inverse of the annihilation operator \cite{mehta1992eigenstates}, namely adds a single photon to the input state at success probability that scales inversely with the number of photons. Furthermore, repeating this process with the outgoing state for a number of iterations larger than the number of photons in the input pulse guarantees successful addition, which is heralded by a toggled state of a following readout photon. 
We then show that the success on $n^{th}$ trial in fact implements what is best described as the $n^{th}$-order bright quantum scissors (BQS) on the input state, which unlike regular quantum scissors (that truncate optical states to contain no more than one photon \cite{pegg1998optical}) produce a state $\ket*{n+}$ that contains \emph{at least} $n$ photons (Fig. \ref{Fig_BQS}a). Beyond the fact that for certain input parameters these bright states approximate Fock states very well, we present a variation of the BQS scheme that ideally results in exact Fock states. Finally, we show that reversing the roles of the output channels and measuring the number of photons in the multiphoton output pulse collapses the train of single-photon pulses from the other output to a polarisation W-state (Fig. \ref{Fig_BQS}b).

The outline of this paper is organised as follows: In Section \ref{sec:framework} we present the theoretical model in which our quantum state evolves. Section \ref{sec:toolbox} is dedicated to presenting and acquiring intuition for SPRINT-based multi-photon processes. In Section \ref{sec:protocol} we introduce the multi-step protocol. Finally, in Section \ref{sec:results} we show how the inverse annihilation operator and the BQS can be employed on the traveling light field and how to produce the aforementioned Fock and W states.

\section{\label{sec:framework}Theorerical Framework}
Consider the cavity-mediated interaction of an optical field with a three-level $\Lambda$ system where each transition is coupled
to one of two orthogonal polarisations; denote them
as the horizontal (H) and vertical (V) polarisations (Fig. \ref{Fig_SPRINT}). Throughout this study we refer to the $\Lambda$ system as an atom, however this is merely an matter of convenience and should not limit the results to a specific physical implementation.
Using the MOU approach, this system can be described by the following Hamiltonian \cite{gea2013photon}: 
\begin{align}
\label{hamiltonian}
\mathcal{H=} -i\hbar g\intop d\omega\frac{\sqrt{\kappa/\pi}}{\kappa-i\omega}(\ket*{e}\bra{g_{h}}\hat{a}_{\omega}+\ket*{e}\bra{g_{v}}\hat{b}_{\omega})e^{-i(\omega+\delta)t}+h.c
\end{align}
where $\kappa$ is the cavity amplitude decay rate which is proportional to the width of its resonance. All the frequencies are relative to the cavity resonance frequency $\Omega_{c}$; the detuning of the atomic transition from the resonance of the cavity is denoted by $\delta=\Omega_{c}-\omega_{a}$ and the detuning
of the actual light field frequency, $\Omega$, from the cavity is
denoted by $\omega=\Omega-\Omega_{c}$. The operators $\hat{a}_{\omega}$
and $\hat{b}_{\omega}$ are the annihilation operators for the H-
and V-mode, respectively. These operators obey the continuum
commutation relations in the frequency-domain $[\hat{a}_{\omega},\hat{a}_{\omega'}^{\dagger}]=[\hat{b}_{\omega},\hat{b}_{\omega'}^{\dagger}]=\delta(\omega-\omega')$.
The parameter $g$ represents the cavity-atom coupling strength where
$2g$ is the rate equal to the single-photon Rabi frequency. 

\begin{figure}[b]
	\centering\includegraphics[width=12cm]{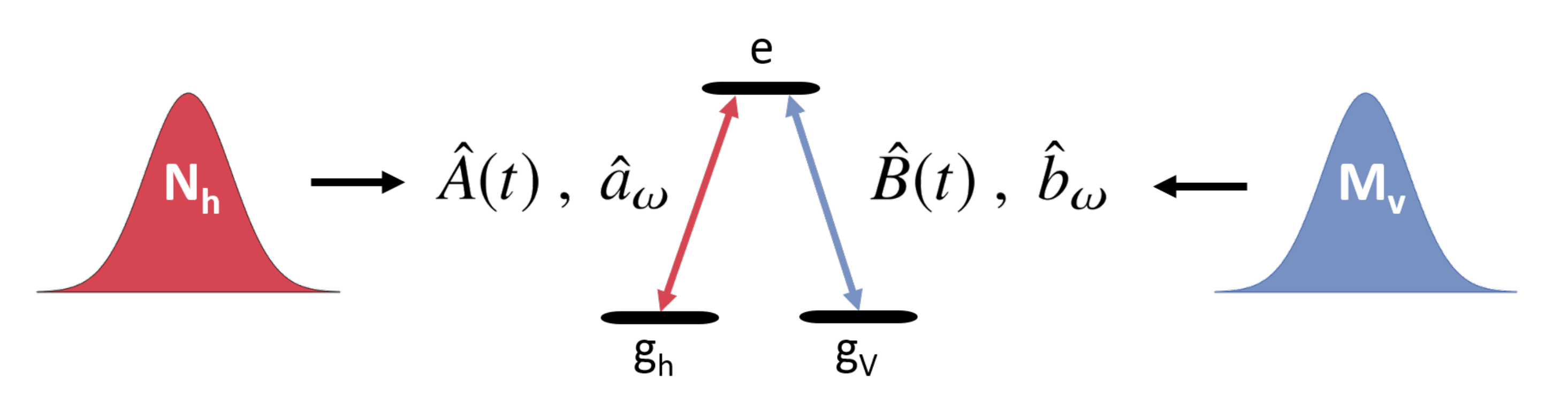}
	\caption{One-dimensional atom. The effective system considered using the MOU approach in the adiabatic limit. Two modes of light, $\hat{a}_{\omega}$ and $\hat{b}_{\omega}$ (or $\hat{A}(t)$ and $\hat{B}(t)$), interact
		with the two transitions of an atom in a $\Lambda$ configuration.}
	\label{MOU}
\end{figure}

Following \cite{gea2013photon}, we work under several conditions.
First, the cavity is on-resonance with the atomic transition, i.e $\delta=0$.
Second, throughout the analytical derivation we assume that cavity losses and free-space spontaneous emission are negligible. Moreover, we assume two adiabatic limits related to $T$, the duration of the pulses we use; $\kappa T\gg1$ and $\Gamma T\gg1$ where $\Gamma=\frac{2g^{2}}{\kappa}$.
In fact, $\Gamma$ is the cavity-enhanced spontaneous emission rate
of the atom to the mode of the cavity. Therefore, in these terms,
the requirement of negligible free-space spontaneous emission translates
to large cooperativity $C\equiv\frac{\Gamma}{\gamma}\gg1$. Under these conditions our system is described effectively by Fig. \ref{MOU}, often referred to as the fast-cavity limit or the one-dimensional atom \cite{turchette1995one}. This space-time
approach has been shown to be equivalent to the well-known ``input-output''
formalism \cite{gardiner1985input,gardiner1993driving,carmichael1993quantum}
when the cavity transmission losses are small enough to allow for
a Lorenzian approximation to the cavity resonance line \cite{gea2013space}.

It is necessary to introduce a few concepts that will help set the
stage for developing the quantum state engineering protocol. As in
\cite{gea2013photon}, we will make use of the field annihliation
operators
\begin{align}
\label{annih}
\hat{A}(t)\equiv & \frac{1}{\sqrt{2\pi}}\int d\omega\hat{a}_{\omega}e^{-i\omega t}\\
\hat{B}(t)\equiv & \frac{1}{\sqrt{2\pi}}\int d\omega\hat{b}_{\omega}e^{-i\omega t}
\end{align}
which can be thought of as the Fourier transform of the frequency
domain operators $\hat{a}_{\omega}$ and $\hat{b}_{\omega}$. It is easy to see
that these obey the continuum commutation relations in the time-domain
$[\hat{A}(t),\hat{A}^{\dagger}(t')]=[\hat{B}(t),\hat{B}^{\dagger}(t')]=\delta(t-t')$.In addition, we can define an N-photon wavepacket in the H-mode in following
manner,
\begin{equation}
\label{fockN}
\ket*{N_{h}}=\frac{1}{\sqrt{N!}}\left(\int dtf(t)\hat{A}^{\dagger}(t)\right)^{N}\ket*{0}
\end{equation}
where $f(t)$ is the pulse-shape of the wavepacket and the state is
normalized for $\int dt|f(t)|^{2}=1$. An N-photon wavepacket in the
V-mode, $\ket*{N_{v}}$, can be described by simply replacing $\hat{A}^{\dagger}$
with $\hat{B}^{\dagger}$ in the expression above. Lastly, we introduce a
state of N photons in the H-mode and a single photon in the V-mode;
this state is time-entangled such that the V-photon is created in
the $k^{th}$ time-slot (where $k\in\left\{ 1,...,N+1\right\} $)

\begin{align}
\label{timeentang}
\ket*{N_{h},\overset{k^{th}}{1_{v}}}\equiv
&\frac{\sqrt{(N+1)!}}{\left(k-1\right)!\left(N-k+1\right)!}\intop_{-\infty}^{\infty}dt\hat{B}^{\dagger}(t)f(t)\times\nonumber \\
&\left[\intop_{-\infty}^{t}dt_{1}\hat{A}^{\dagger}(t_{1})f(t_{1})\right]^{k-1}\left[\intop_{t}^{\infty}dt_{2}\hat{A}^{\dagger}(t_{2})f(t_{2})\right]^{N-k+1}\ket*{0} 
\end{align}
In other words, as opposed to a the product state $\ket*{N_{h}}\otimes\ket*{1_{v}}$
where the time-ordering of the photons is unknown, in state (\ref{timeentang})
we can be certain that the photon in the V-mode was created after
exactly $\left(k-1\right)$ photons in the H-mode. 

\section{\label{sec:toolbox}SPRINT-based Toolbox}
SPRINT, previously presented in \cite{rosenblum2011photon,rosenblum2017analysis}
using the input-output formalism, can be expressed in terms of the MOU approach.
The evolution of initial state $\ket*{1_{h},0_{v},g_{h}}$ under Hamiltonian (\ref{hamiltonian}) is in fact a special case of the photon subtraction
described in \cite{gea2013photon}; following the interaction with
the atom, the initial state $\ket*{N_{h},0_{v},g_{h}}$ is transformed to the final state $\ket*{N-1_{h},\overset{1^{st}}{1_{v}},g_{v}}$.
Substituting $N=1$ in this result provides us with the desired effect, the initial H-photon is converted to a V-photon while the atom toggles from state $\ket*{g_h}$ to $\ket*{g_v}$:
\begin{equation}
\label{subtraction}
\ket*{1_{h},0_{v},g_{h}}\to-\ket*{0_{h},1_{v},g_{v}}
\end{equation}

Utilising SPRINT as a building block we can assemble a toolbox, which
consists of the evolution of two specific states. The multi-step protocol in the next section leans heavily on these two processes;
effective time-shifting and deterministic photon addition described
in Eq. (\ref{addnshift}a) and (\ref{addnshift}b), respectively.
\begin{subequations}
	\label{addnshift}
	\begin{align}
	\ket*{N_{h},\overset{k^{th}}{1_{v}},g_{v}}\to & \ket*{N_{h},\overset{(k+1)^{th}}{1_{v}},g_{v}}\\
	\ket*{N_{h},\overset{(N+1)^{th}}{1_{v}},g_{v}}\to & -\ket*{N+1_{h},0_{v},g_{h}}
	\end{align}
\end{subequations}
One can obtain these processes by solving the time-dependent Schr\"{o}dinger equations associated with the evolution of the corresponding initial states in the same manner as in \cite{gea2013photon}. Instead of presenting the cumbersome derivation of these processes, we introduce a simple intuition for these results using SPRINT. Generally, we can picture a multi-photon process
in the following way; in the adiabatic limit where the pulse is very
long compared to the inverse of the cavity-enhanced decay rate, the
probability of having two photons time-spaced by less than $\frac{1}{\Gamma}$
is negligible. Hence, we can conclude that each photon within the
pulse interacts with the atom-cavity separately. When each photon
reaches the atom-cavity, one in two may happen; if the atom is in
the ground state matching the mode of the photon ($\ket*{1_{v},g_{v}}$
or $\ket*{1_{h},g_{h}}$), the resulting photon is emitted in
the other mode and the atom toggles to the other ground state,
in accordance with SPRINT. In the other case, where the atom is in
a ground state not matching the mode of the photon ($\ket*{1_{h},g_{v}}$
or $\ket*{1_{v},g_{h}}$), no interaction will occur since the optical field
is not coupled to the relevant transition.

Now it is easy to get intuition for Eq. (\ref{addnshift}a). Since we start with the
atom in $\ket*{g_{v}}$, the first $(k-1)$ H-photons do not interact with the atom. The $k^{th}$ photon is in the V-mode,
therefore it experiences SPRINT which results in the atom
toggling to $\ket*{g_{h}}$ and an H-photon emitted. Then for the $(k+1)^{th}$
H-photon we have SPRINT again (since the atom is now in $\ket*{g_{h}})$,a V-photon is emitted leaving the atom in $\ket*{g_{v}}$.
The remaining $(N-k)$ H-photons in the pulse have no interaction
with the atom. Consequently, the resulting state is a V-photon in
the $(k+1)^{th}$ time-position and all the rest $N$ photons in the
H-mode. Overall, this process describes effective time-shifting
of the V-photon; from the $k^{th}$ time-slot to the $\left(k+1\right)^{th}$
time-slot.

An exception to the above considerations is the case where $k=N+1$,
i.e the V-photon arrives last as noted in the initial state of Eq. (\ref{addnshift}b). Similarly, the first $N$ H-photons do not interact with the
atom and the $(N+1)^{th}$ V-photon experiences SPRINT, toggling
the atom to $\ket*{g_{h}}$ and emitting an H-photon. Since it was
the last photon we do not have another SPRINT as in the previous case.
Therefore we are left with $(N+1)$ H-photons and the atom in $\ket*{g_{h}}$,
which is the final state described in Eq. (\ref{addnshift}b). As a consequence, we
get that the single photon in the V-photon is added deterministically
to the $N$ photons in the H-mode.

In general, we do not have time-entangled initial states at our disposal
such as those used in the time-shifting and deterministic addition
processes. Therefore, we present a mathematical identity (\ref{identity}) that
links the product state $\ket*{N_{h},1_{v}}$ to these time-entangled
states. Basically, it describes this product state as an equal superposition
of the time-entangled states representing all the different $\left(N+1\right)$
time-ordering of the photons.
\begin{equation}
\label{identity}
\ket*{N_{h},1_{v}}=\frac{1}{\sqrt{N+1}}\left(\ket*{N_{h},\overset{1^{st}}{1_{v}}}+...+\ket*{N_{h},\overset{\left(N+1\right)^{th}}{1_{v}}}\right)
\end{equation}

\section{\label{sec:protocol}Multi-step Protocol}
\begin{figure}[t]
	\centering\includegraphics[width=10cm]{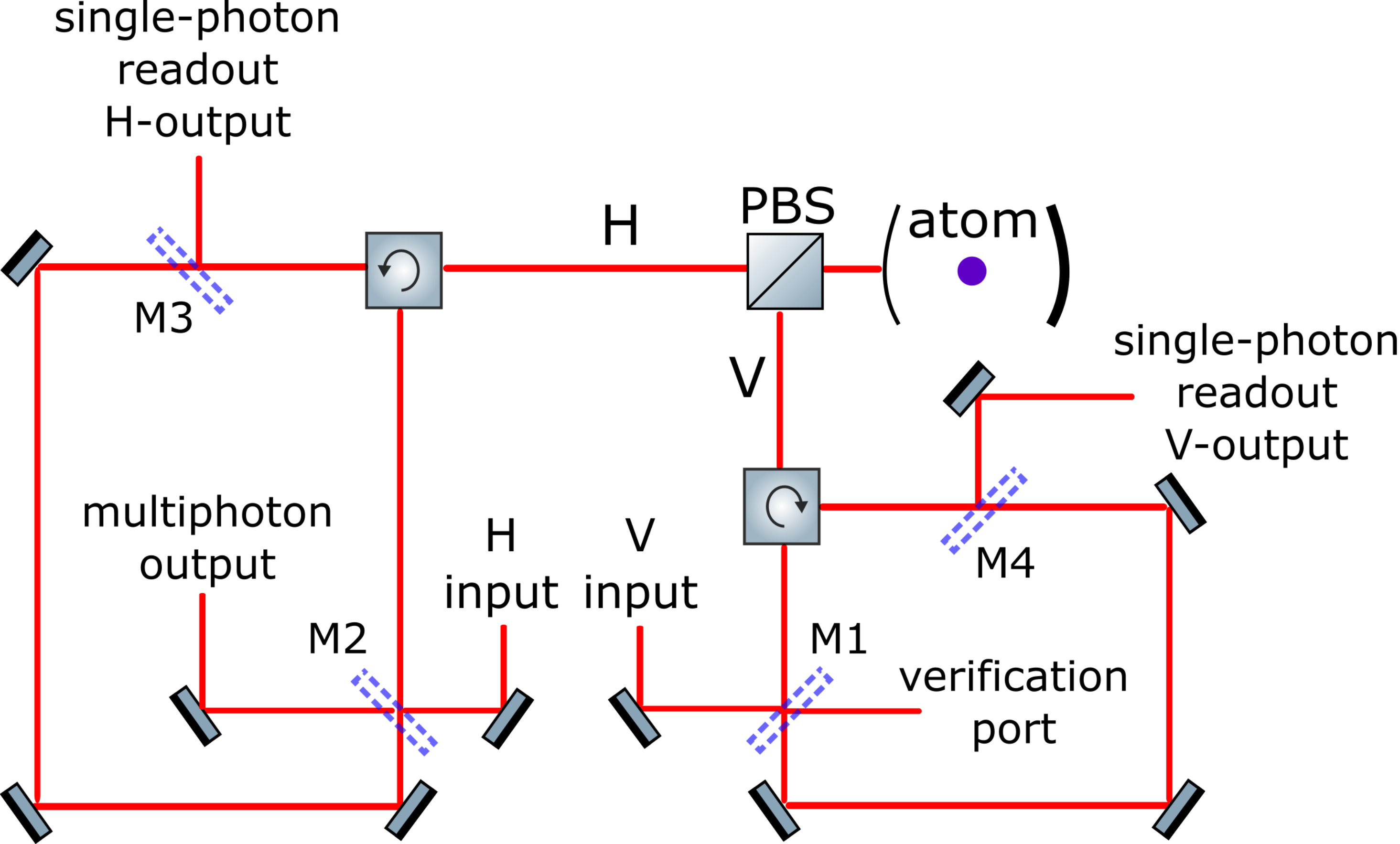}
	\caption{Optical setup suitable for the implementation of the BQS protocol.
		H-input and V-input are the sources for the pulses in the two modes. Switchable mirrors M1-M4 are used to repeatedly alternate between directing the readout photons to their respective outputs and rerouting the multiphoton state back into the cavity. Upon measuring a photon in the ''single-photon readout V-output'' on the $n^{th}$ iteration of the protocol, M2 is turned ON and an $\ket*{n+}$ state is measured in ''multiphoton output''. On the other hand, when heralding on vacuum in the verification port and $M$ photons in the multiphoton output, a W state is generated at the combination of single-photon readout H- and V-output.}
	\label{opticalsetup1}
\end{figure}
\begin{figure}[t]
\centering\includegraphics[width=11cm]{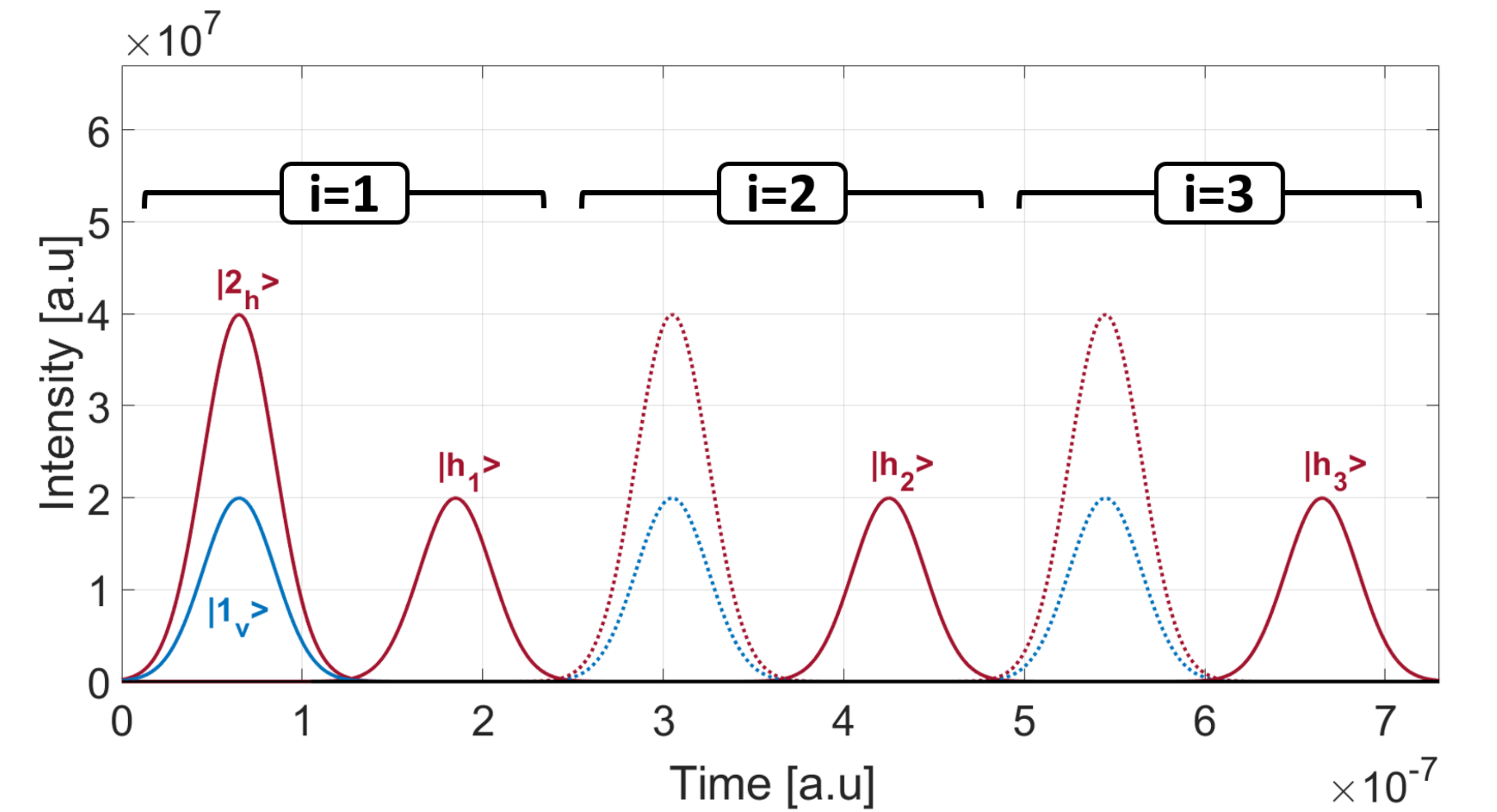}
	\caption{Experimental pulse sequence for $\ket*{\phi_{h}}=\ket{2_{h}}$. The H- and V-modes are represented by red and blue
		respectively. Solid lines refer to pulses we send actively from H-
		and V-input while dotted lines represent those redirected back into
		the cavity. The iteration number appears above the relevant pulses.}
	\label{pulses}
\end{figure}

Based on processes (\ref{subtraction}) and (\ref{addnshift}), we have constructed an iterative protocol for QSE. The first step of the protocol involves interacting the atom initialised in $\ket*{g_{v}}$ with a multiphoton state comprised of two simultaneous pulses; a general H-polarised state and a single V-photon, $\ket*{\phi_{h},1_{v}}$. Following the interaction, the pulses reflected off the cavity are rerouted back into
the system by switchable mirrors (realised using Pockels cells) keeping the H- and V-modes the same (Fig. \ref{opticalsetup1}). While these
pulses are being rerouted, we send an additional single H-photon
in order to reinitialise the atom to $\ket*{g_{v}}$ using SPRINT
(\ref{subtraction}). As a result, either an H- or a V-photon can be emitted, depending
on the final state of the atom after the initial pulses have completed
the interaction. Subsequently, the rerouted multiphoton state interacts with the atom once again.
This sequence is repeated as depicted in Fig. \ref{pulses}; we refer to
a single iteration of the protocol as interacting the multiphoton state (or its evolutions)
with the atom followed by reinitialising the atom. The train of single photons resulting from the reinitialisation photons is henceforth referred to as ''readout photons'' and denoted $\ket*{h_{i}}$
or $\ket*{v_{i}}$ where the subscript indicates the number of iteration. The readout photons are directed to the single-photon readout output (either H or V) by switchable mirrors (M3 and M4), and thus seperated from the multiphoton state. Finally, upon proper heralding on the readout channel we can realise the inverse annihilation and bright scissors operation on the multiphoton state. On the other hand, heralding on the multiphoton output channel and the verification port (using M1 and M2), we can generate polarisation W states in the readout photons. These are discussed in detail in section \ref{sec:results}.

In order to get intuition for the iterative protocol we examine
the evolution of the initial state $\ket*{1_{h},1_{v},g_{v}}$ in Eq.
(\ref{evolution}). For convenience, we denote the interaction of the multiphoton state
with the atom as $\overset{atom}{\longrightarrow}$ and the $j^{th}$
reinitialisation of the atom using an H-photon by $\overset{h_{j}}{\longrightarrow}$.
Using identity (\ref{identity}) and the tools provided in Eq. (\ref{subtraction}) and (\ref{addnshift}) it is simple
to follow the evolution of the state throughout the protocol.
\begin{center}
	\begin{align}
	\label{evolution}
	&\ket*{1_{h},1_{v},g_{v}}=
	\frac{1}{\sqrt{2}}(\ket*{1_{h},\overset{2^{nd}}{1_{v}}}+\ket*{1_{h},\overset{1^{st}}{1_{v}}})\otimes\ket*{g_v}\\
	& \overset{atom}{\longrightarrow}\frac{1}{\sqrt{2}}(-\ket*{2_{h},0_{v},g_{h}}+\ket*{1_{h},\overset{2^{nd}}{1_{v}},g_{v}})\overset{h_{1}}{\longrightarrow}\frac{1}{\sqrt{2}}(\ket*{2_{h},0_{v},g_{v}}\otimes\ket*{v_{1}}+\ket*{1_{h},\overset{2^{nd}}{1_{v}},g_{v}}\otimes\ket*{h_{1}})\nonumber \\
	& \overset{atom}{\longrightarrow}\frac{1}{\sqrt{2}}\left(\ket*{2_{h},0_{v},g_{v}}\otimes\ket*{v_{1}}-\ket*{2_{h},0_{v},g_{h}}\otimes\ket*{h_{1}}\right)\overset{h_{2}}{\longrightarrow}\frac{1}{\sqrt{2}}\ket*{2_{h},0_{v},g_{v}}\otimes\left(\ket*{v_{1},h_{2}}+\ket*{h_{1},v_{2}}\right)\nonumber\\
	& \longrightarrow\underbrace{......}_{2\leq k+1\:iterations}\longrightarrow\ket*{2_{h},0_{v},g_{v}}\otimes\frac{1}{\sqrt{2}}\left(\ket*{v_{1},h_{2},h_{3},...,h_{k+1}}+\ket*{h_{1},v_{2},h_{3},...,h_{k+1}}\right) \nonumber
	\end{align}
\end{center}

It is constructive to think of the protocol in terms of photon addition.
The state $\ket*{N_{h},1_{v}}$ has an equal probability
of having each of the $N+1$ possible time-orderings of the V-photon (Eq. (\ref{identity})). For the time-ordering in which the V-photon
is last, the resulting field state after interaction with the atom
is $\ket*{N+1_{h},0_{v}}$ (Eq. (\ref{addnshift}a)) i.e. the V-photon was added to the H-mode. As for the other possible time-orderings, the time-position of the V-photon will move one slot to a later
time (Eq. (\ref{addnshift}b)). Therefore, $N+1$ repeated attempts of photon addition with the initial $\ket*{N_{h},1_{v}}$ state guarantee that the V-photon is added to
the H-mode. In our iterative scheme, the additional H-photon we send
serves two goals; first, it reinitialises the atom to $\ket*{g_{v}}$
allowing repeated addition attempts. Second, since a successful addition
leaves the atom in $\ket*{g_{h}}$, the following emitted readout photon tells
us whether the addition was successful (V-photon) or not (H-photon).
Hence, through entanglement of our state to the readout photons,
we have information about when (at which iteration or attempt) did a
successful addition occur. With this in mind, we can generalize Eq. (\ref{evolution})
to an initial $\ket*{N_{h},1_{v}}$ state and look at the outcome of
the protocol after $\left(k+1\right)$ iterations
\begin{center}
	\begin{equation}
	\label{evolution2}
	\ket*{N_{h},1_{v},g_{v}}\mathrel{\mathop{\longrightarrow}^{\mathrm{(k+1)}}_{\mathrm{iterations}}}\begin{cases}
	&\textbf{for }\mathbf{k\geq N:}\\ &\ket*{N+1_{h},0_{v},g_{v}}\otimes\frac{1}{\sqrt{N+1}}\bigg(\ket*{v_{1},h_{2},h_{3},...,h_{k+1}}+\ket*{h_{1},v_{2},h_{3},...,h_{k+1}}\\
	& +...+\ket*{h_{1},...,h_{N},v_{N+1},h_{N+2},...h_{k+1}}\bigg)\\
	&\textbf{for }\mathbf{k\leq N-1:}\\ &\ket*{N+1_{h},0_{v},g_{v}}\otimes\frac{1}{\sqrt{N+1}}\bigg(\ket*{v_{1},h_{2},h_{3},...,h_{k+1}}+\ket*{h_{1},v_{2},h_{3},...,h_{k+1}}\\
	& +...+\ket*{h_{1},...,h_{k},v_{k+1}}\bigg)+\\
	& +\frac{1}{\sqrt{N+1}}\underbrace{\left(\ket*{N_{h},\overset{(k+2)^{th}}{1_{v}},g_{v}}+...+\ket*{N_{h},\overset{(N+1)^{th}}{1_{v}},g_{v}}\right)}_{(N-k)\:terms}\otimes\ket*{h_{1},...,h_{k+1}}
	\end{cases}
	\end{equation}
	\par\end{center}

We can now determine the outcome of any initial state in the H-mode
and a single-photon in the V-mode. Expanding the arbitrary state in
the H-mode using the Fock basis we can write the initial state
\begin{equation}
\label{psiini}
\ket*{\psi_{initial}}=\ket*{\phi_{h},1_{v}}=\sum_{N=0}^{\infty}C_{N}\ket*{N_{h},1_{v}}
\end{equation}
Using Eq. (\ref{evolution2}) we can get the resulting state after $\left(k+1\right)$
iterations of the scheme
\begin{align}
\label{psifinal}
& \ket*{\psi_{final}}=\nonumber\\
& \ket*{v_{1},h_{2},h_{3},...,h_{k+1}}\otimes\left(\sum_{N=0}^{\infty}\frac{C_{N}}{\sqrt{N+1}}\ket*{N+1_{h}}\right)\ket*{0_{v},g_{v}}+\nonumber \\
& \ket*{h_{1},v_{2},h_{3},...,h_{k+1}}\otimes\left(\sum_{N=1}^{\infty}\frac{C_{N}}{\sqrt{N+1}}\ket*{N+1_{h}}\right)\ket*{0_{v},g_{v}}+\nonumber \\
& \vdots \\
& \ket*{h_{1},...,h_{k},v_{k+1}}\otimes\left(\sum_{N=k}^{\infty}\frac{C_{N}}{\sqrt{N+1}}\ket*{N+1_{h}}\right)\ket*{0_{v},g_{v}}+\nonumber \\
& \ket*{h_{1},...,h_{k+1}}\otimes\sum_{N=k+1}^{\infty}\frac{C_{N}}{\sqrt{N+1}}\Bigg(\ket*{N_{h},\overset{(k+2)^{th}}{1_{v}}}+\nonumber \\
& +...+\ket*{N_{h},\overset{(N+1)^{th}}{1_{v}}}\Bigg)\ket*{g_{v}}\nonumber
\end{align}
Heralding differently will allow us to engineer quantum states and
implement various operations.

\section{\label{sec:results}Results}
\subsection{Inverse Annihilation}
Since the annihilation operator has an eigenvalue of zero for $a\ket*{0}=0$,
we cannot find an operator $\hat{O}$ such that $\hat{O}\hat{a}=I$
. On the other hand, we can find $\hat{O}$ which satisfies $\hat{a}\hat{O}=I$,
this is known as the inverse annihilation operator $a^{-1}$ \cite{mehta1992eigenstates},
\begin{equation}
\label{invannih}
aa^{-1}=I\:\:;\:\:a^{-1}a=I-\ket*{0}\bra{0}
\end{equation}
In the Fock basis representation it has the form
\begin{equation}
\label{invannihfock}
a^{-1}=\sum_{n=0}^{\infty}\frac{1}{\sqrt{n+1}}\ket*{n+1}\bra{n}
\end{equation}

The operation of the inverse annihilation can be achieved using only
a single step of the protocol presented above. Looking at Eq. (\ref{psifinal})
we can see that if we herald on $\ket*{v_{1}}$, this is exactly the
operation we get for the initial H-mode state $\ket*{\phi_{h}}=\sum_{N=0}^{\infty}C_{N}\ket*{N_{h}}$.
Since we herald on $\ket*{v_{1}}$ we need just one iteration of the protocol,
i.e $k=0$.
\begin{align}
\label{invannihherald}
\braket*{v_{1}}{\psi_{final}}&= \left(\sum_{N=0}^{\infty}\frac{C_{N}}{\sqrt{N+1}}\ket*{N+1_{h}}\right)\ket*{0_{v},g_{v}}\nonumber\\
&\overset{\star}{\to}\sum_{N=0}^{\infty}\frac{C_{N}}{\sqrt{N+1}}\ket*{N+1_{h}}\\
&=a^{-1}\sum_{N=0}^{\infty}C_{N}\ket*{N_{h}}=a^{-1}\ket*{\phi_{h}}\nonumber
\end{align}
where in $\star$ we trace over the atom and the V-mode. This effect is actually described in \cite{gea2013photon}
as a probabilistic photon addition but in fact, since it changes the photon
number statistics, it does not function as the addition operator $\hat{A}=\sum_{n=0}^{\infty}\ket*{n+1}\bra{n}$ (as was performed with phonons of a trapped ion \cite{um2016phonon})
but rather as the inverse annihilation $a^{-1}$. 

\begin{figure}[t]
	\centering\includegraphics[width=10cm]{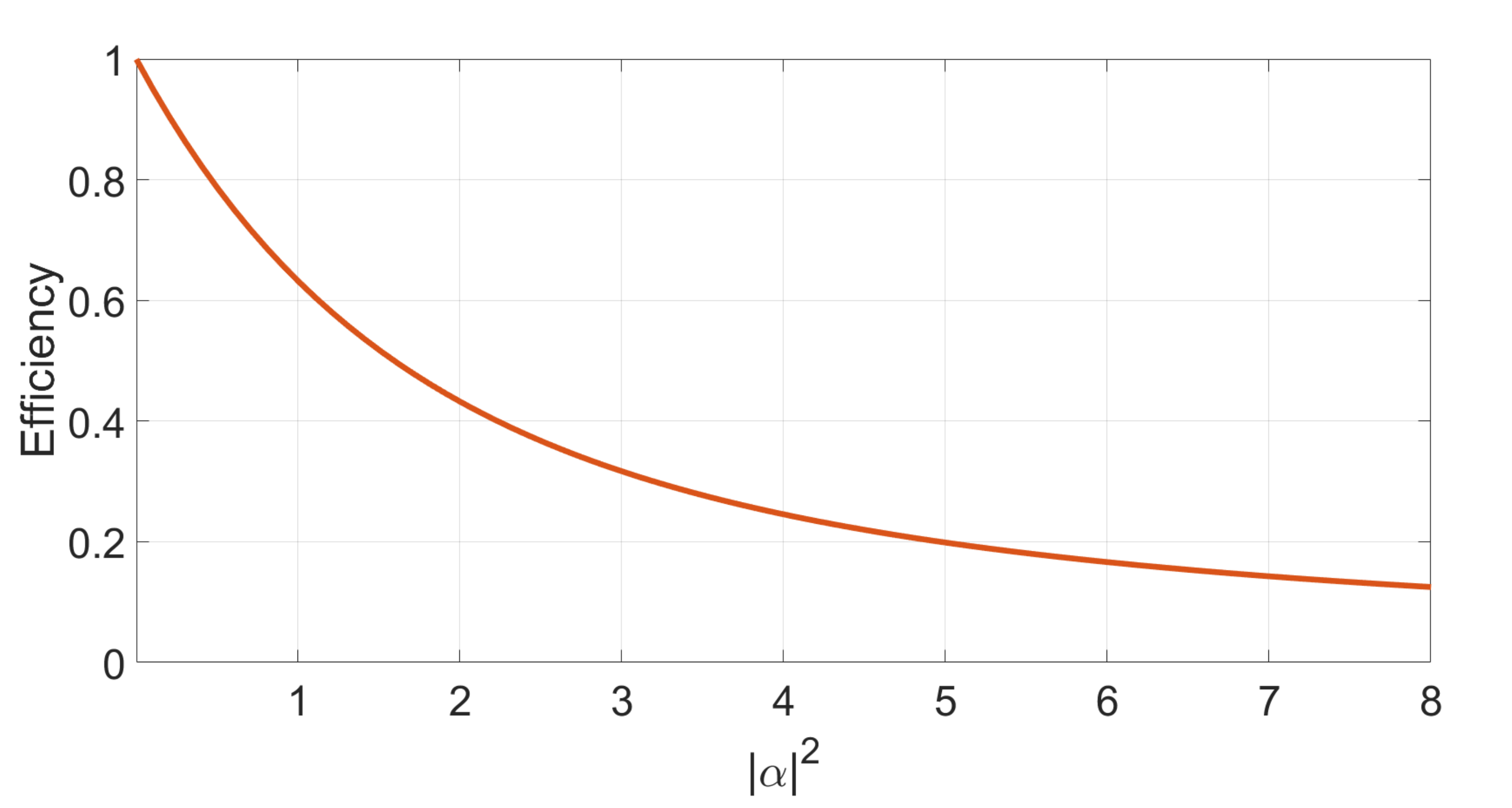}
	\caption{Efficiency of $a^{-1}$ as a function of $\left|\alpha\right|^{2}$,
		the average number of photons in the initial coherent state. }
	\label{invanniheff}
\end{figure}

Fidelity and efficiency are used to characterize the quality of a
process. Fidelity is a measure to quantify accuracy, it is the overlap
between the final state of the process and the ideal, desired state.
Efficiency, on the other hand, is the probability to obtain this final
state by the end of the process. Upon heralding on $\ket*{v_{1}}$,
the process is of unit fidelity and the efficiency of this process is
given by
\begin{equation}
\label{eta1}
\eta_{1}=\sum_{N=0}^{\infty}\frac{\left|C_{N}\right|^{2}}{N+1}
\end{equation}
For an initial coherent state $\ket*{\alpha}$ in the H-mode  
we get the efficiency of the inverse annihilation operator described in Fig. \ref{invanniheff}.

\subsection{\label{secBQS}Bright Quantum Scissors}
\begin{figure}[t]
	\centering\includegraphics[width=11cm]{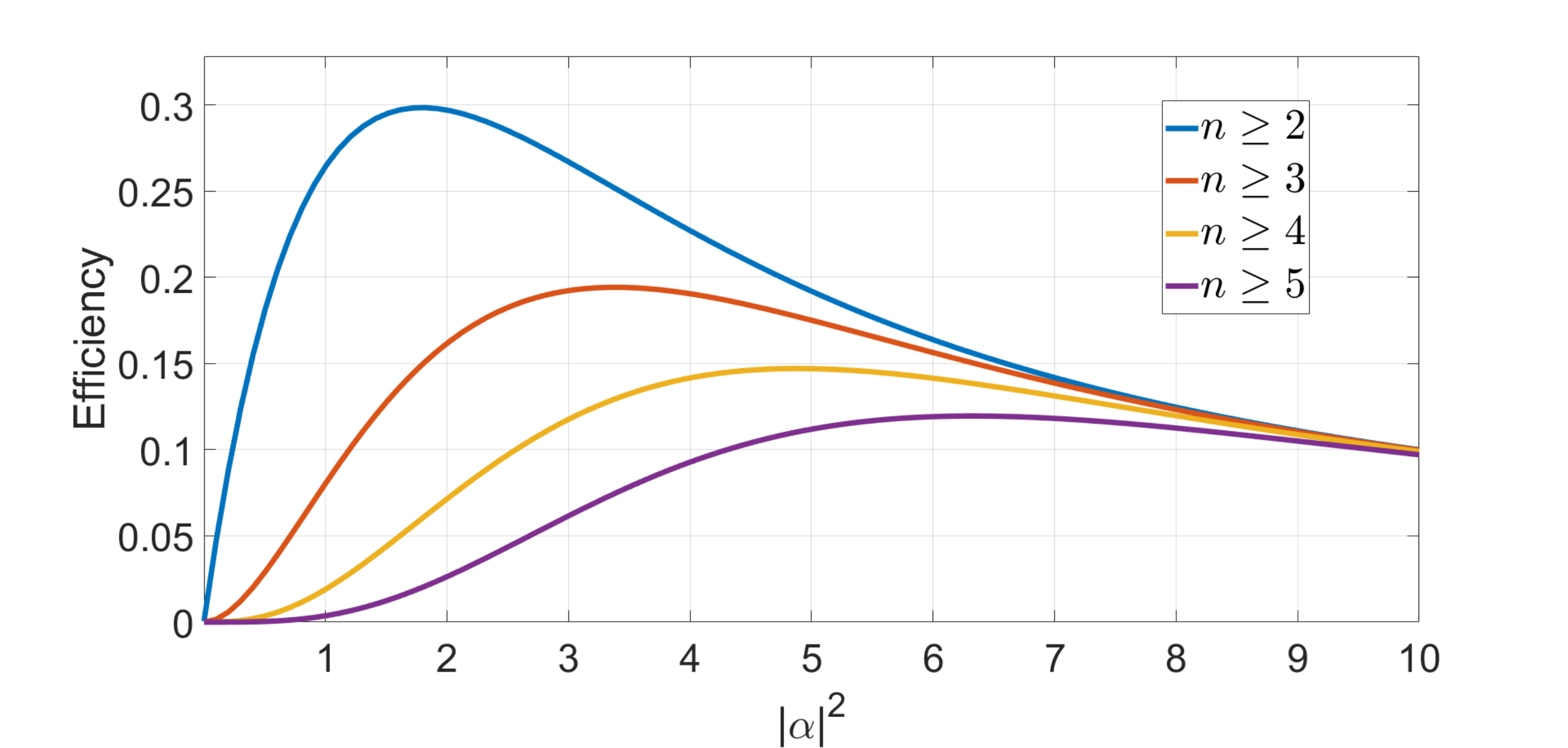}
	\caption{$n^{th}$-order BQS. Efficiency as a function of the average number of
		photons in the initial coherent state, $\left|\alpha\right|^{2}$.
		The resulting state $\ket*{n+}$ is guaranteed to have more than $n$ photons.}
	\label{brightscissorseff}
\end{figure}

One may characterize a quantum state $\ket*{\psi}$ using its photon-number distribution defined by the probabilities $P\left(N\right)=|\braket*{N}{\psi}|^2$. The $k^{th}$-order BQS operation truncates any input quantum state such that the modified state has at least $k$ photons,
i.e $P\left(N<k\right)=0$. This is in some sense complementary
to the well-known quantum scissors introduced in \cite{pegg1998optical}, which leaves only the vacuum and one-photon components of the quantum state.
Looking at Eq. (\ref{psifinal}), we see that heralding on $\ket*{v_{k}}$ ensures the operation of the $k^{th}$-order BQS.
\begin{align}
\label{brightscissors}
\braket*{v_{k}}{\psi_{final}}&=
\underbrace{\ket*{h_{1},h_{2},...,h_{k+1}}}_{trace\;out}\otimes\left(\sum_{N=k-1}^{\infty}\frac{C_{N}}{\sqrt{N+1}}\ket*{N+1_{h}}\right)\otimes\underbrace{\ket*{0_{v},g_{v}}}_{trace\:out}\nonumber \\
&\rightarrow\mathcal{N}\sum_{N=k}^{\infty}\frac{C_{N-1}}{\sqrt{N}}\ket*{N_{h}}\equiv\ket*{k+}
\end{align}
where $\mathcal{N}$ is a normalization factor.
This resulting state is a highly non-classical since the probability $P(N<k)$ vanishes \cite{lutkenhaus1995nonclassical}. The process
is of unit fidelity and the efficiency of the $k^{th}$-order BQS is given by
\begin{equation}
\label{eta2}
\eta_{2}=\sum_{N=k}^{\infty}\frac{\left|C_{N-1}\right|^{2}}{N}
\end{equation}
For an input coherent state in the H-mode we get the efficiency presented in Fig. \ref{brightscissorseff}.

The BQS operation (including its $1^{st}$-order interpretation as the inverse annihilation) can be made deterministic by choosing the number of iterations in accordance with the photon-number distribution of the input state. As previously discussed, for an initial $\ket*{N_h,1_v}$, addition is guaranteed after $(k+1)$ or more iterations and the resulting readout V-photon tells us at which iteration did it occur. Therefore, by choosing the number of iterations such that $P(N\ge k+1)$ of the general input state is negligible (\ref{condition_deterministicBQS}), we can be certain that BQS was performed and the order of its operation is indicated by the readout V-photon. 
\begin{equation}
P(N\ge k+1)=\sum_{N=k+1}^{\infty}|C_N|^2\ll1
\label{condition_deterministicBQS}
\end{equation}
As can be seen from Eq. (\ref{psifinal}), the probability of BQS acting on the input state after $(k+1)$ iterations is given by
\begin{equation}
\eta_{BQS} = 1-\sum_{N=k+1}^{\infty}\frac{(N-k)\left|C_{N}\right|^{2}}{N+1}
\label{eta_bqs}
\end{equation} 
In the case where the number of iterations and the photon-number distributon of the input state maintain condition (\ref{condition_deterministicBQS}), the sum in Eq. (\ref{eta_bqs}) vanishes, making the operation of the BQS deterministic.

\begin{figure}[b]
	\centering\includegraphics[width=11cm]{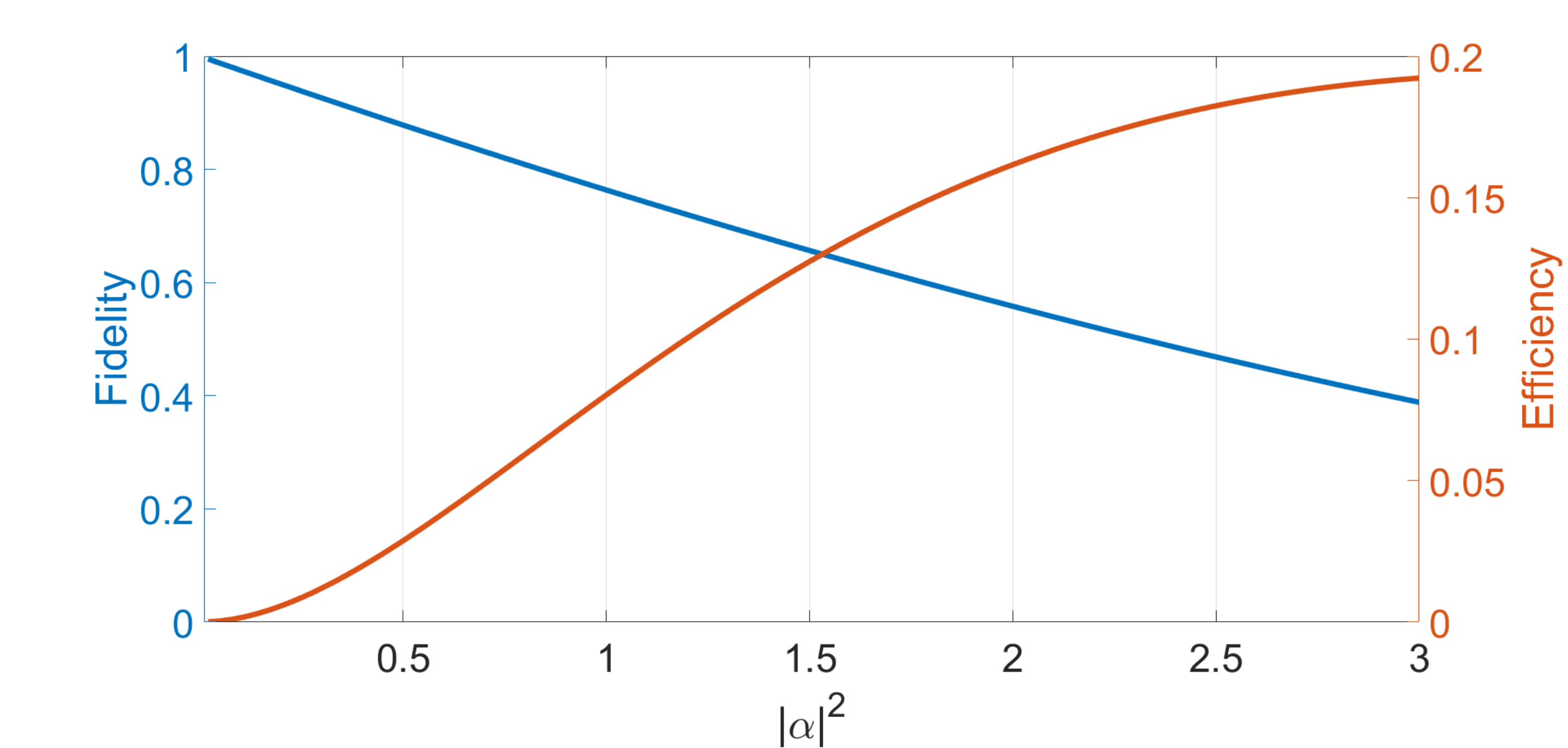}
	\caption{Generation of a Fock state $\ket*{3}$ using the $3^{rd}$-order BQS. Fidelity and efficiency as a function of the average number of photons in the
	coherent state.}
	\label{brightscissorsfock}
\end{figure}

BQS can also be used to generate Fock states from coherent state input (Fig. \ref{brightscissorsfock}) by choosing
$\left|\alpha\right|^{2}$ small enough such that the probability
of $\ket*{k}$ in the resulting state (\ref{brightscissors}) will be much larger than
that of $\ket*{k+1}$ and higher components. The relation between these
probabilities will determine the fidelity of the Fock state. Clearly,
there is a trade-off between the efficiency and the fidelity of the
process; choosing a lower average number of photons in the coherent
state results in higher fidelity since the probabilities of $\ket*{k+1}$
or higher components decrease relative to the probability of $\ket*{k}$.
On the other hand, this low number of photons also leads to a low efficiency. A better scheme for producing Fock states is described in subsection (\ref{fockgen}).

The BQS described in Eq. (\ref{brightscissors}) alters the ratio between the amplitudes of the remaining number states. If we wish to ''cut the tail''
of the photon-number distribution while also keeping the ratios of the initial state (\ref{psiini}) the
same, we can operate on our initial state with the BQS
followed by the annihilation operator (typically using a high-transmittivity
beam splitter \cite{wenger2004non}). This results in
\begin{equation}
\label{neutralbrightscissors}
\hat{a}\mathcal{N}\sum_{N=k}^{\infty}\frac{C_{N}}{\sqrt{N+1}}\ket*{N+1_{h}}=\mathcal{N}\sum_{N=k}^{\infty}C_{N}\ket*{N_{h}}
\end{equation}
which is equivalent to the operator
\begin{equation}
\hat{O}=\hat{I}-\sum_{n=0}^{k-1}\ket*{n}\bra{n}
\end{equation}
acting on the H-polarised initial state. Hence, at the price of
an additional iteration and a decrease in efficiency due to the annihilation
process, we can get a neutral-BQS operation that maintains the ratio between probability amplitudes of the initial state.

\subsection{\label{fockgen}Fock State Generation}
\begin{figure}[t]
	\centering\includegraphics[width=10cm]{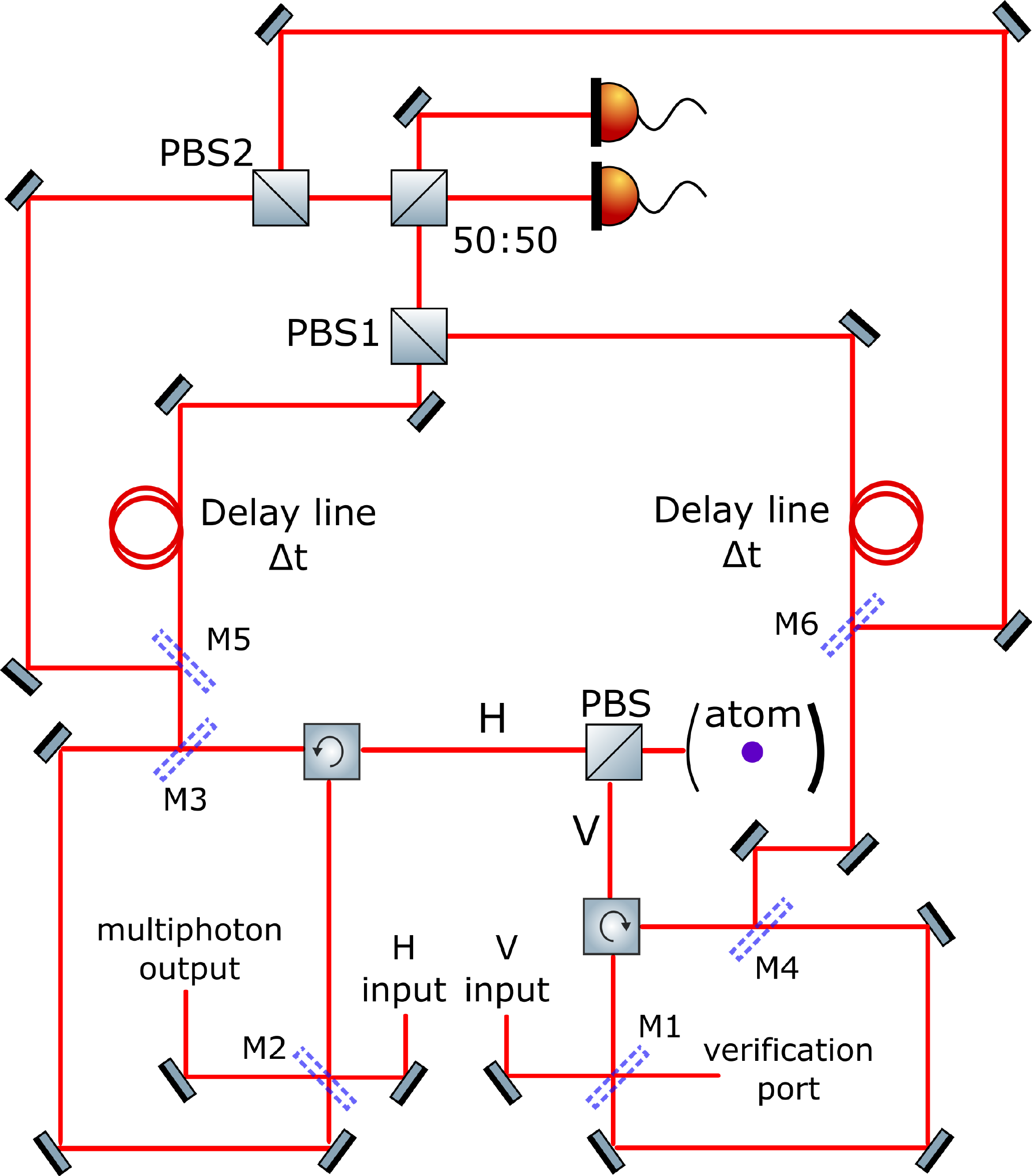}
	\caption{Optical setup for Fock state generation via Bell state measurement. First, we direct the $k^{th}$
		readout photon (either H or V) to a delay line (mirrors M5 and M6 are OFF). When the $(k+1)^{th}$ readout photon leaves the
		cavity we turn M5 and M6 ON in order to direct it to the path with no time delay. The delay time is
		set such that both the $k^{th}$ and the $\left(k+1\right)^{th}$ readout
		photons enter the 50:50 beam splitter simultaneously. Coincident detections
		at the output of the 50:50 beam splitter guarantee that our state
		has collapsed on the antisymmetric Bell state \cite{braunstein1995measurement}.}
	\label{opticalsetup2}
\end{figure}

Using an interference-based measurement of two consecutive readout photons, we are able to generate Fock states with unit fidelity. For this purpose we
must alter the readout output ports in order to realise a Bell state measurement (Fig. \ref{opticalsetup2}). 

Consider an entangled state in the form 
\begin{equation}
\ket*{\chi}=\ket*{v_{k}}\ket*{h_{k+1}}\ket*{\psi_{1}}+\ket*{h_{k}}\ket*{v_{k+1}}\ket*{\psi_{2}}
\end{equation}
After passing through the optical setup we have four possible modes
for the two readout photons reaching the 50:50 beam splitter simultaneously;
two different incoming ports denoted by the subscript and two different
polarisations, vertical (V) and horizontal (H). Therefore, we can
rewrite the state as
\begin{equation}
\ket*{\chi}=\ket*{V_{1}H_{2}}\ket*{\psi_{1}}+\ket*{H_{1}V_{2}}\ket*{\psi_{2}}
\end{equation}
Then, heralding on coincident detections in the two photodetectors
we collapse on the antisymmetric Bell state \cite{braunstein1995measurement}
\begin{equation}
\pm\frac{1}{\sqrt{2}}\bra{\psi^{(-)}}=\pm\frac{1}{\sqrt{2}}\left(\bra{V_{1}H_{2}}-\bra{H_{1}V_{2}}\right)
\end{equation}
Therefore,
\begin{equation}
\pm\frac{1}{\sqrt{2}}\braket*{\psi^{(-)}}{\chi}=\pm\frac{1}{\sqrt{2}}\left(\ket*{\psi_{1}}-\ket*{\psi_{2}}\right)
\end{equation}

Implementing this measurement on our final state in Eq. (\ref{psifinal}) for the
$k^{th}$ and the $(k+1)^{th}$ outgoing readout photons we get
(ignoring the overall sign)
\begin{align}
& \frac{1}{\sqrt{2}}\left(\bra{v_{k},h_{k+1}}-\bra{h_{k},v_{k+1}}\right)\ket*{\psi_{final}}=\nonumber \\
& =\frac{1}{\sqrt{2}}\underbrace{\ket*{h_{1},...,h_{k-1}}\otimes\ket*{0_{v},g_{v}}}_{trace\:out}\otimes\left(\sum_{N=k-1}^{\infty}\frac{C_{N}}{\sqrt{N+1}}\ket*{N+1_{h}}-\sum_{N=k}^{\infty}\frac{C_{N}}{\sqrt{N+1}}\ket*{N+1_{h}}\right)\nonumber \\
& =\frac{1}{\sqrt{2}}\frac{C_{k-1}}{\sqrt{k}}\ket*{k_{h}}
\end{align}
This means that heralding on coincident detections of the $k^{th}$
and $(k+1)^{th}$ readout photons we get a Fock state of $\ket*{k}$
with unit fidelity and an efficiency be given by 
\begin{equation}
\eta_{3}=\frac{\left|C_{k-1}\right|^{2}}{2k}
\end{equation}
We can understand it intuitively as interferring two BQS
operations; one providing an output state containing more than $k$
photons and the other a state with more than $\left(k+1\right)$ photons.
Then, if the interference is with a minus sign we get a telescoping
sum leaving just the Fock state of $\ket*{k}$.

\begin{figure}[t]
	\centering\includegraphics[width=10cm]{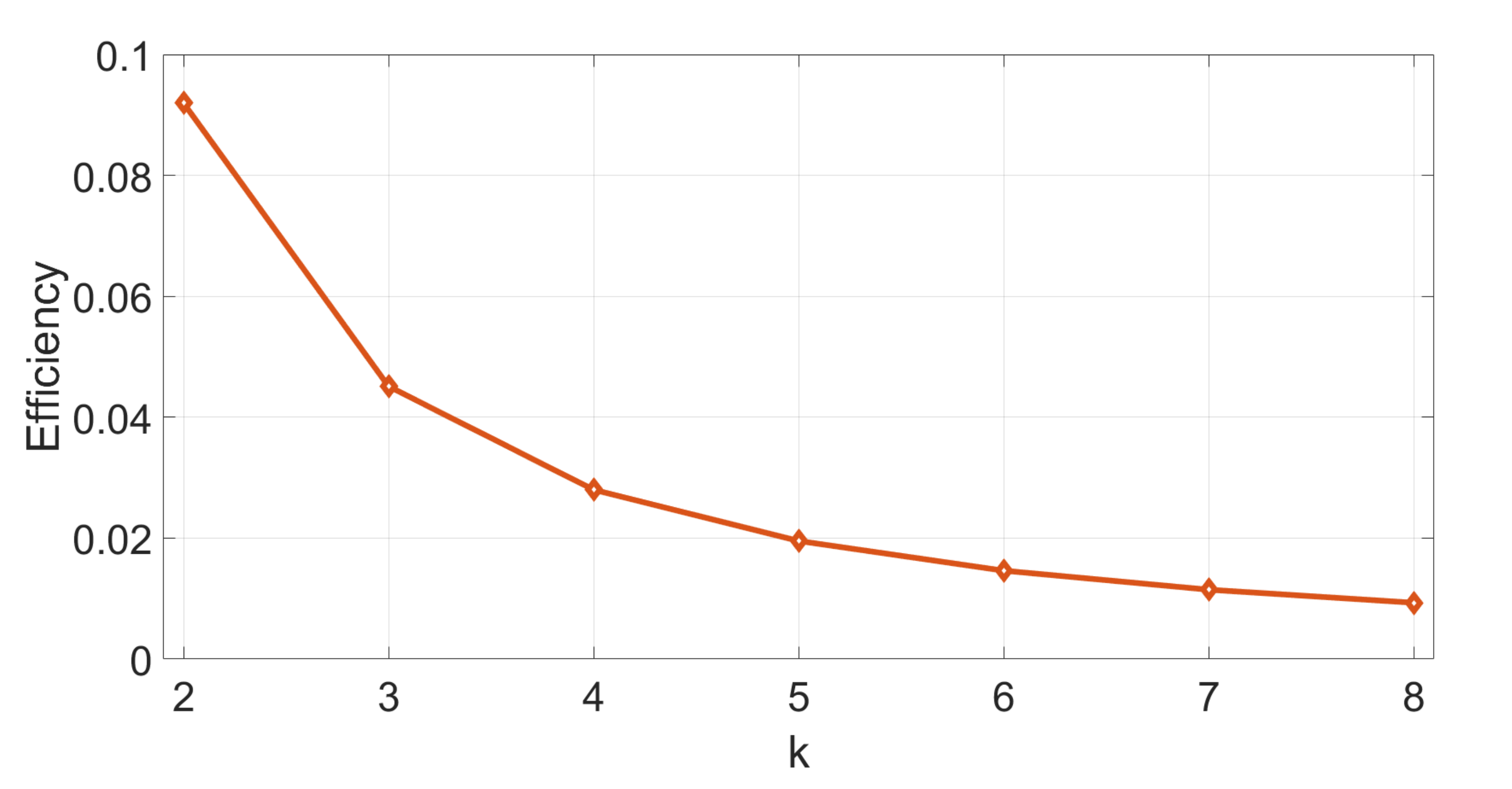}
	\caption{Efficiency of generating a Fock state $\ket*{k}$ by using
		an optimal coherent state input (see text).}
	\label{fockeff}
\end{figure}

Using an input coherent state $\ket*{\alpha}$ we can optimize the efficiency of generating Fock $\ket*{k}$
by choosing the average number of photons $\left|\alpha\right|^{2}=k-1$
such that $\left|C_{k-1}\right|^{2}$ is maximised. Fig. \ref{fockeff} presents
this optimal efficiency for various Fock states.

\subsection{W State Generation}
An n-qubit W state in the polarisation basis is defined for $n\geq3$ bellow. 
\begin{equation}
\label{Wdef} \ket*{W_{n}}=\frac{1}{\sqrt{n}}\left(\ket*{VHH...H}+\ket*{HVH...H}+...+\ket*{HH...HV}\right)
\end{equation}

In order to generate W states using the BQS protocol we reverse roles; heralding is performed on the multiphoton output and the resulting W state is comprised of the readout photons. It is then constructive to rearrange the terms in the final state of the protocol (\ref{psifinal}) to the following form
\begin{align}
\label{psifinal2}
&\ket*{\psi_{final}}=\nonumber\\ &\sum_{N=0}^{k-1}\frac{C_{N}}{\sqrt{N+1}}\ket*{N+1_{h},0_{v},g_{v}}\otimes\nonumber\\
&\bigg(\ket*{v_{1},h_{2},h_{3},...,h_{k+1}}+\ket*{h_{1},v_{2},h_{3},...,h_{k+1}}+...+\ket*{h_{1},...,h_{N},v_{N+1},h_{N+2},...h_{k+1}}\bigg)\nonumber \\
&+\sum_{N=k}^{\infty}\frac{C_{N}}{\sqrt{N+1}}\ket*{N+1_{h},0_{v},g_{v}}\otimes\nonumber\\
&\bigg(\ket*{v_{1},h_{2},h_{3},...,h_{k+1}}+\ket*{h_{1},v_{2},h_{3},...,h_{k+1}}+......+\ket*{h_{1},...,h_{k},v_{k+1}}\bigg)\nonumber \\
&+\sum_{N=k+1}^{\infty}\frac{C_{N}}{\sqrt{N+1}}\left(\ket*{N_{h},\overset{(k+2)^{th}}{1_{v}},g_{v}}+...+\ket*{N_{h},\overset{(N+1)^{th}}{1_{v}},g_{v}}\right)\otimes\ket*{h_{1},...,h_{k+1}}
\end{align}
Following the operation of the protocol for 3 or more iterations, we deflect the multiphoton state to the multiphoton output and verification port using mirrors M1 and M2 (see Fig. \ref{opticalsetup1}). If one measures the multiphoton output in the state of $\ket*{M_{h},0_{v}}$ (for $M\geq3$) then the remaining readout photons collapse to

\begin{center}
	\begin{equation}
	\label{heraldW}
	\braket*{M_{h},0_{v}}{\psi_{final}}=\frac{C_{M-1}}{\sqrt{M}}\ket*{g_{v}}\otimes\begin{cases}
	&\textbf{for }\mathbf{3\leq M\leq k:}\\ &\ket*{v_{1},h_{2},h_{3},...,h_{k+1}}+\ket*{h_{1},v_{2},h_{3},...,h_{k+1}}+\\
	&+...+\ket*{h_{1},...,h_{M-1},v_{M},h_{M+1},...h_{k+1}}\\
	&\textbf{for }\mathbf{M\geq k+1}:\\ 
	&\ket*{v_{1},h_{2},h_{3},...,h_{k+1}}+\ket*{h_{1},v_{2},h_{3},...,h_{k+1}}+\\
	& +...+\ket*{h_{1},...,h_{k},v_{k+1}}
	\end{cases}
	\end{equation}
	\par\end{center}
Tracing over the state of the atom, renormalizing and using definition (\ref{Wdef}) results in
\begin{center}
	\begin{equation}
	\label{Wstates}
	\braket*{M_{h},0_{v}}{\psi_{final}}=
	\begin{cases}
	\ket*{W_M}\otimes\ket*{h_{M+1},...,h_{k+1}}\;\;&\text{for }3\leq M\leq k\\
	\ket*{W_{k+1}}&\text{for }M\geq k+1
	\end{cases}
	\end{equation}
	\par\end{center}

\begin{figure}[b]
	\centering\includegraphics[width=10cm,trim={1.5cm 0 1.5cm 0},clip]{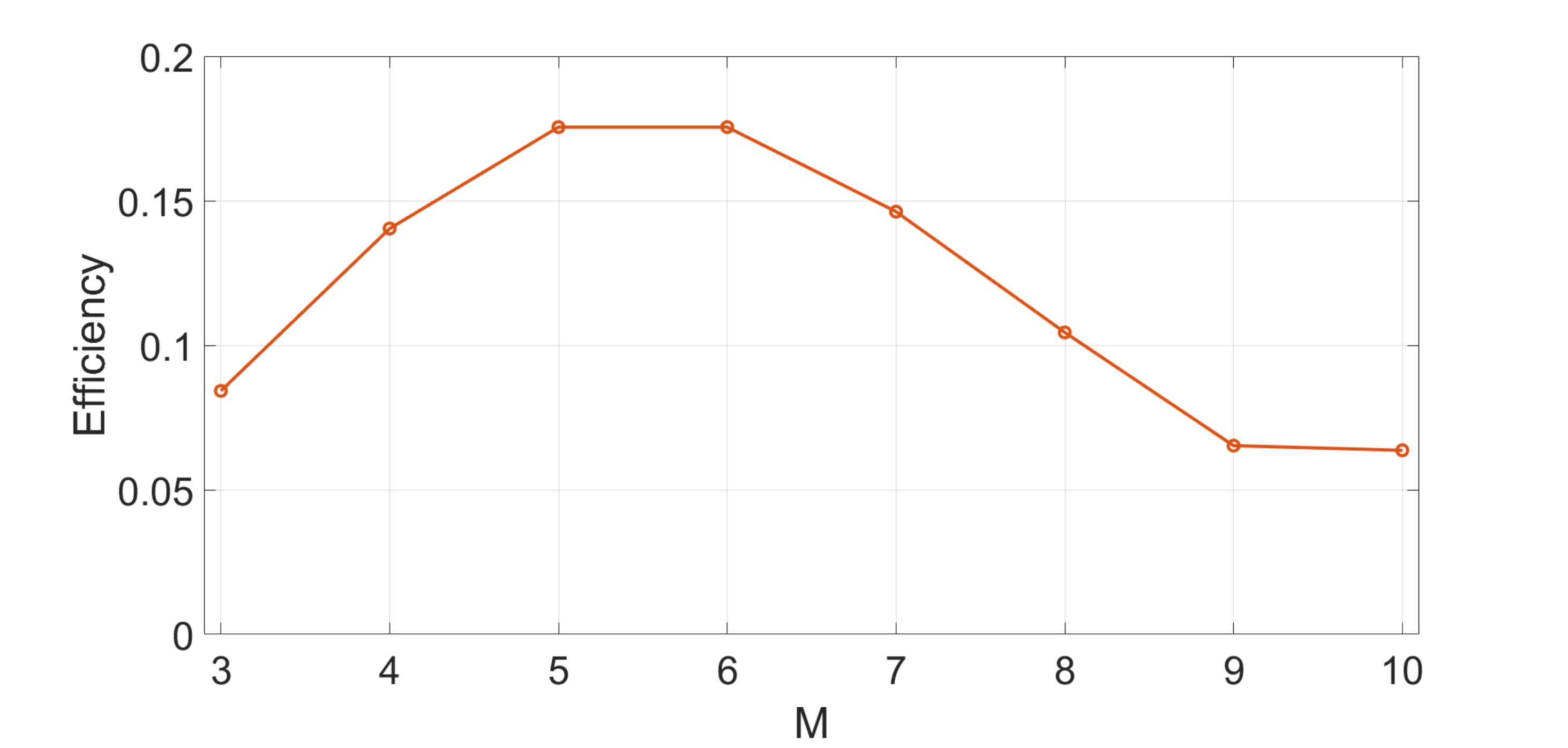}
	\caption{Efficiency for generating $W_M$ states by using an initial coherent state with average photon number of 5 and applying the protocol for 10 iterations.}
	\label{Wstatesuccess}
\end{figure}

We can think of the generation of W states in the following manner; whenever we find vacuum in the verification port and $M\geq3$ H-photons the multiphoton output, we are guaranteed to have a unit-fidelity W state manifested in the time-seperated readout photons. In the case where $M$ is greater than or equal to the number of iterations we get a W state with the number of qubits equal to the number of iterations. On the other hand, when $M$ is smaller than the number of iterations, we get a W state of size $M$ and additional $(k-M)$ H-photons that may be ignored for any practical purpose.

As an example, examine the action of 10 iterations of the protocol on an initial coherent state with average photon number of 5 in the H-mode. The success probability for generating a $\ket*{W_{M}}$ state is depicted in Fig. \ref{Wstatesuccess}. Notice that the sum of these probabilities approaches unity ($\sim96\%$), therefore the production of any W state is near-deterministic. This is always the case when the number of iterarions of the protocol is larger than the number of photons in the multiphoton input state, as in the BQS operation (see section \ref{secBQS}). In addition, since the success probability of $\ket*{W_{k+1}}$ is comprised of all the contributions of more than $(k+1)$ signal photons in the H-mode, there is a clear enhancement of the success probability for $\ket*{W_{10}}$.

\section{\label{sec:feas}Feasibility}
There are a few issues that need addressing in terms of experimental feasibility. The scheme assumes an on-demand single-photon source for the initial V-mode and for the train of H-photons, as well as unit single-photon detection probability for indication and heralding. It also assumes high cooperativity, namely negligible interaction with optical modes that are not Purcell-enhanced by the cavity. In all these, significant progress has been made in recent years. The field of all-optical quantum information processing has motivated major efforts both towards the attainment of deterministic single-photon sources, quantum-dot based \cite{somaschi2016near,he2017deterministic,daveau2017efficient} and others \cite{kaneda2015time,jeantet2016widely}, and towards efficient superconducting single-photon detectors \cite{marsili2013detecting}. Novel waveguide and cavity technology, photonic band-gap in particular, reach cooperativities approaching $10^2$ \cite{arcari2014near}. However, the most deleterious issue is optical loss. In order to get intuition on the effects of loss on the fidelity of this scheme, consider the production of a Fock state of
$\ket*{3}$ using the bright scissors operation (Fig. \ref{brightscissorsfock}) with an
initial H-mode coherent pulse of $\left\langle N\right\rangle =0.02$.
Upon correct heralding, i.e measuring $\ket*{h_{1},h_{2},v_{3}}$, it is most
probable that the outgoing state has evolved from the initial $\ket*{2_{h},\overset{1^{st}}{1_{v}}}$
component of the coherent state. This is so since a lower number of
photons cannot result in a $\ket*{v_{3}}$ photon (successful addition
in the third attempt), while higher number of photons is less probable
by several orders of magnitude due to the low average number of photons.
In addition, any other time-ordering of the $\ket*{2_{h},1_{v}}$ state
where the V-photon is not first, will not lead to a $\ket*{v_{3}}$
photon. Then let us examine the evolution of $\ket*{2_{h},\overset{1^{st}}{1_{v}}}$
through the three repetitions of the protocol; a loss of a photon
or more during the first repetition will result in one of the following: $\ket*{2_{h},0_{v}}$,$\ket*{1_{h},\overset{2^{nd}}{1_{v}}}$,$\ket*{1_{h},0_{v}}$,$\ket*{0_{h},1_{v}}$ and $\ket*{0}$. None of those states can result in $\ket*{h_{1},h_{2},v_{3}}$
since they will either toggle the atom on the next step producing
$\ket*{v_{2}}$ or not toggle the atom at all leading to no readout
V-photon during the entire protocol. Hence, the loss on the first
repetition will not affect the fidelity and we can consider the ideal
state $\ket*{2_{h},\overset{2^{nd}}{1_{v}}}$ as the only one contributing
to the next steps. In contrast, during the second and third repetitions,
a loss of a photon could still generate the correct heralding but the protocol will not result in the final Fock state $\ket*{3}$. Hence,
the fidelity is governed by a factor of $\left(1-L\right)^{6}$ (where
$L$ is the loss of the cavity) signifying that no photon was lost
in any of the six SPRINT interactions of these two repetitions. This
power law, which appears for other cases as well, amounts to a significant
decrease in fidelity and poses an obstacle for the experimental implementation
of such a multi-step protocol. Nonetheless, the on-going technological development in manufacturing high-Q and low-loss
optical resonators \cite{ji2017ultra,pfeiffer2017coupling,yang2018bridging}, is expected to bring the demonstration of W and Fock states with moderate number of photons to within reach in the near future.

\section{\label{sec:summ}Summary}
In this work we described a protocol for optical QSE that performs the BQS operation on any input quantum state. The protocol is based on repeated SPRINT iterations of the input state together with single-photon pulses, carried out by a single $\Lambda$ system in a single-sided cavity in the Purcell regime. We note that strong coupling is not necessary for SPRINT, as well as for most photon-atom gates \cite{borne2019deterministic}. The special case of a single iteration of the BQS protocol realises the inverse annihilation operator. Multiple iterations can be used to deterministically generate a single pulse in a bright quantum state $\ket*{n+}$ that has at least $n$ photons, or a train of single photon pulses in a $W_{n}$ state. In both cases the specific value of $n$ is indicated by a measurement at the other output port, and the probabilities for different values of $n$ are determined by the initial input quantum state (e.g. a coherent state $\ket*{\alpha}$). While at certain input parameters the state $\ket*{n+}$ approximates well the Fock state $\ket*{n}$, a variation of the protocol can be used to produce heralded exact Fock states. The main vulnerability of the protocol is linear loss, which hampers its scaling-up to a large number of photons. Accordingly our efforts are now aimed at adding more heralding mechanisms into the protocol, to allow maintaining fidelity of the generated states at the expense of lower efficiency. Nonetheless, with the advancements of technologies for efficient generation and detection of single photons, together with the on-going efforts towards coupling quantum emitters such as atoms, ions, quantum dots and spin-systems to low-loss, high quality waveguides and resonators \cite{ji2017ultra,pfeiffer2017coupling,yang2018bridging,davanco2017heterogeneous,goban2014atom}, this protocol could serve as a versatile building-block for QSE in quantum communication, distributed quantum information processing and all-optical quantum computing.

\section*{Funding}
BD acknowledges support from the Israeli Science Foundation, the Minerva Foundation and the Crown Photonics Center. BD is also supported by a research grant from Charlene A. Haroche and Mr. and Mrs. Bruce Winston.

MSK is supported by the KIST Institutional Program (2E26680-18-P025), the Samsung GRO project, the Royal Society and the EU BlinQ project.
BD and MSK are supported by the Weizmann-UK joint research program.

\section*{Acknowledgments}
This research was made possible in part by the historic generosity of the Harold Perlman family.

\bibliography{QSE_lit}{}
\end{document}